\documentclass[12pt]{article}
\usepackage{graphicx}
\usepackage[cp1251]{inputenc}
\usepackage{epsfig}
\usepackage[english]{babel}

\date{}

\begin{document}
\setcounter{page}{1}
\pagestyle{plain}

\title{\bf{Magnetism of an adatom on biased AA-stacked bilayer graphene}}

\author{Yawar Mohammadi$^1$\thanks{Corresponding author. Tel./fax: +98 831 427
4569, Tel: +98 831 427 4569, PO Box: 67144-15111. E-mail address:
yawar.mohammadi@gmail.com} , Rostam Moradian$^{2,3}$}
\maketitle{\centerline{$^1$Department of Physics, Islamic Azad
University, Kermanshah Branch, Kermanshah, Iran}
\maketitle{\centerline{$^2$Department of Physics, Razi University,
Kermanshah, Iran} \maketitle{\centerline{$^3$Nano Science and Nano
Technology Research Center, Razi University, Kermanshah,Iran }

\begin{abstract}
We study magnetism of an adatom adsorbed on AA-stacked bilayer
graphene (BLG) in both unbiased and biased cases, using the
Anderson impurity model. We find different magnetic phase diagrams
for the adatom, depending on the energy level of the adatom, which
varies from the magnetic phase diagram of adatom in normal metals
to that in graphene. This is due to the individual energy
dependence of the density of states (DOS) of AA-stacked BLG and
anomalous broadening of the adatom energy level. Furthermore we
investigate the effect of a bias voltage on DOS of AA-stacked and
show that the magnetization of the adatom can be controlled by
applying the bias voltage. This allows for possibility of using
AA-stacked BLG in spintronic devices.
\end{abstract}

%{\it \emph{PACS}}: \emph{74.20.-z, 74.20.Fg}

\vspace{0.5cm}


%\newpage
\section{Introduction}
\label{sec:1}

Graphene is a single layer of carbon atoms arranged in a hexagonal
lattice structure. Low-energy quasiparticles in graphene have
linear dispersion relation and behave as massless Dirac
particles~\cite{Novoselov}. This special dispersion relation leads
to many unusual properties have not been observed before graphene
isolation~\cite{Castro Neto,Peres}. Properties of few-layer
graphene materials depend on the number of layers and the stacking
order~\cite{Castro Neto}. For example BLG in AB stacking has
gapless quadratic dispersion relation and shows
properties~\cite{Das Sarma,McCann} which are different from
graphene and also from the ordinary two dimensional electron gas.
Also AA-stacked BLG, a new stable stacking order of
graphene~\cite{Liu,Borysiuk}, has properties which are different
from those of SLG and BLG. In the AA-stacked BLG each sublattice
of the top layer is located directly above the same one of the
bottom layer. The AA-stacked BLG has special band structure
composed of two hole-dopped and electron-dopped graphene-like
bands~\cite{Ando,Hsu}. Its band structure includes two decouple
doped SLG bands which are shifted relative to each other. Due to
this especial band structure the AA-stacked BLG shows attractive
properties~\cite{Ando,Hsu,Prada,Tabert,Chiu,Brey} which have not
been observed in the other two dimensional materials.

In two previous decade, some successful experimental
efforts~\cite{Eigler,Brar} to control the position of an impurity
adsorbed on a two-dimensional open surface by scanning tunnelling
microscope have been reported. These motivated several groups to
consider magnetization of adatoms adsorbed on
garphene~\cite{Uchoa,Cornaglia,Mao,Li,Wehling1,Wehling2} and
AB-stacked BLG~\cite{Ding,Killi,Sun,Mohammadi}. They showed that
the magnetization of the adatom adsorbed on graphene and BLG is
more possible than the magnetization of an adatom embedded in
normal metals. Also they found that it can be controlled by an
electric field via back gate~\cite{Uchoa}. For graphene, these
features have been verified by an experimental research recently
~\cite{Nair}. The AA-stacked BLG, which is a new open surface two
dimensional material with especial low-energy DOS~\cite{Ando,Hsu},
can be appealing to consider the adatom magnetization (Note that
DOS plays key role in the magnetization of the adatom).
Furthermore in this paper we show that the DOS of AA-stacked BLG
can be manipulated by a vertical electric field (bias voltage).
Hence one can control the magnetization of the adatom. This
feature allows for high possible application of the AA-stacked BLG
in the spintronics devices. In this paper, motivated by these
facts, we consider the magnetization of an adatom adsorbed on
biased AA-stacked BLG surface.

This paper is organized as follows. In Sec. II we present model
Hamiltonian and details of our calculation. Also we consider the
effect of the bias voltage on the DOS of AA-stacked BLG which has
key effect on the magnetization of the adatom. Sec. III is devoted
to our results and discussion. First we consider necessary
conditions for the magnetization of the adatom in the unbiased
case. Then we focus on the effect of the bias voltage on the
magnetization of the adatom. Finally the paper is ended with the
summary and conclusions.

\section{Model Hamiltonian}
\label{sec:2}

We consider magnetization of an adatom, with a inner localized
orbital, which is adsorbed on a sublattice of a AA-stacked BLG
lattice. The Hamiltonian of this system can be written as
\begin{eqnarray}
H_{T}=H_{BLG}+H_{ad}+H_{V} ,\label{eq:01}
\end{eqnarray}
where $H_{BLG}$ and $H_{ad}$  are the Hamiltonian of the pure
biased AA-stacked BLG lattice and the Hamiltonian of the adatom
respectively. Also the hybridization of the localized orbital of
the adatom with the conduction sea of biased AA-stacked BLG is
reflected in $H_{V}$.

The momentum-dependent Hamiltonian of the biased AA-stacked BLG in
the nearest-neighbor approximation is given by (we use units such
that $\hbar=1$)
\begin{eqnarray}
H_{BLG}=&-&
t\sum_{m=1}^{2}\sum_{\mathbf{k}\sigma}[\phi(\mathbf{k})a_{m\mathbf{k}\sigma}^{\dag}
b_{m\mathbf{k}\sigma}+
\phi^{\ast}(\mathbf{k})b_{m\mathbf{k}\sigma}^{\dag}a_{m\mathbf{k}\sigma}] \nonumber \\
 &-& V \sum_{m=1}^{2}\sum_{\mathbf{k}\sigma}(-1)^{m}[a_{m\mathbf{k}\sigma}^{\dag}
a_{m\mathbf{k}\sigma}+
b_{m\mathbf{k}\sigma}^{\dag}b_{m\mathbf{k}\sigma}] \nonumber \\
&+&
\gamma\sum_{\mathbf{k}\sigma}[a_{1\mathbf{k}\sigma}^{\dag}b_{2\mathbf{k}\sigma}+
b_{2\mathbf{k}\sigma}^{\dag}a_{1\mathbf{k}\sigma}],\label{eq:03}
\end{eqnarray}
where $a_{m \mathbf{k}\sigma}^{\dag}(a_{m \mathbf{k}\sigma})$
creates (annihilates) an electron with spin $\sigma$ at $A$
sublattice in m-th layer. $t\sim3$ eV and $\gamma\sim0.2$
eV\cite{Xu,Lobato} present the nearest neighbor intralayer and the
interlayer  hopping energies respectively and the applied bias
voltage is equal to 2V. In eq. (\ref{eq:05})
$\phi(\mathbf{k})=\sum_{i=1}^{^{3}}e^{i\mathbf{k}.\mathbf{d}_{i}}$
where $\mathbf{d}_{i}$ are are the nearest neighbor vectors which
have been shown in fig. \ref{f.01}. The corresponding energy bands
are
\begin{eqnarray}
\varepsilon_{\lambda,s}=s\sqrt{\gamma^{2}+V^{2}}+\lambda\phi(\mathbf{k})
,\label{eq:04}
\end{eqnarray}
where s refers to $bonding/antibonding$ label and $\lambda=\pm1$
are the electron-like and hole-like band indexes.  One can expand
$|\phi(\mathbf{k})|$ around Dirac points ($\mathbf{K}$ or
$\mathbf{K}^{'}$) for $|\mathbf{q}|\ll|\mathbf{K}|$ (where
$\mathbf{k}=\mathbf{q}+\mathbf{K}$) to obtain the low energy bands
of biased AA-stacked BLG. In this limit
$|\phi(\mathbf{k})|=v_{F}q$ where $v_{F}=3ta/2$ is Fermi velocity.
Plots of the low energy bands for three values of bias voltage,
V=-$\gamma$, V=0.0 and V=+$\gamma$ have been shown in fig.
\ref{f.02}(a). We see that in the biased AA-stacked BLG the s=+
and s=- bands are shifted by $+\sqrt{\gamma^{2}+V^{2}}/\gamma$
($-\sqrt{\gamma^{2}+V^{2}}/\gamma$) with respect to the same band
in the unbiased case. Local DOS at anyone of equivalent
sublattices in the top layer of biased AA-stacked BLG is
\begin{eqnarray}
N(\omega)=\frac{1}{2D^{2}}[\frac{\Pi+V}{\Pi}|\omega+\Pi|+\frac{\Pi-V}{\Pi}|\omega-\Pi|]
\theta(D-|\omega|),\label{eq:05}
\end{eqnarray}
where $\Pi=\sqrt{\gamma^{2}+V^{2}}$, $D$ is the high-energy cutoff
of AA-stacked BLG bandwidth and $\theta(x)$ is the step function.
Not that to obtain local DOS at sublattices of the bottom layer it
is enough to replace V with -V. In fig. \ref{f.02}(b) we have
plotted the local DOS at the top layer for three values of bias
voltage, V=-$\gamma$, V=0.0 and V=+$\gamma$. As it is clear from
eq. (\ref{eq:05}) and also from fig. \ref{f.02}(b) one can control
the local DOS around the adatom level (so the magnetization of the
adatom) via a bias voltage.

The adatom Hamiltonian, $H_{ad}$, is given by
\begin{eqnarray}
H_{ad}=(\varepsilon_{0}+V)\sum_{\sigma}f_{\sigma}^{\dag}f_{\sigma}+
Un_{\uparrow}n_{\downarrow},\label{eq:05}
\end{eqnarray}
where ($\varepsilon_{0}$+V) is the energy level of the adatom
shifted by the bias voltage (Let us suppose that the adatom is
located near the top layer so they have same electric potential.)
and $U$ is the Coulomb energy for double occupancy of the adatom
level. $n_{\sigma}=f_{\sigma}^{\dag}f_{\sigma}$ is the occupation
number of the adatom level and $f_{\sigma}^{\dag}(f_{\sigma})$
operator creates (annihilates ) an electron with spin $\sigma$ at
the adatom level. By using the mean field approximation, the
two-body coulomb interaction term can be decoupled to the
single-body interactions as $\sum_{\sigma}\langle
n_{-\sigma}\rangle f_{\sigma}^{\dag}f_{\sigma}-\langle
n_{\uparrow}\rangle \langle n_{\downarrow}\rangle$. So we can
rewrite the adatom Hamiltonian as $
\sum_{\sigma}\varepsilon_{\sigma}f_{\sigma}^{\dag}f_{\sigma},\label{eq:11}
$ where $\varepsilon_{\sigma}=\varepsilon_{0}-U\langle
n_{-\sigma}\rangle$ is renormalized energy of the adatom level.

The hybridization of the localized orbital of the adatom with the
conduction sea of AA-stacked BLG is given by
\begin{eqnarray}
H_{V}=\frac{V_{f}}{\sqrt{N}}\sum_{\sigma \mathbf{k}
}(f_{\sigma}^{\dag}a_{1 \mathbf{k} \sigma}+a_{1 \mathbf{k}
\sigma}^{\dag}f_{\sigma}),\label{eq:06}
\end{eqnarray}
where $V_{f}$ is the hybridization strength and the Hamiltonian
has been written for an adatom adsorbed on an $A$ sublattice at
top layer and N is the number of $A$ sublattices.

To determine the magnetization of the adatom one must obtain the
occupation number of the spin up and down of the adatom level. If
$n^{ad}_{\uparrow}\neq n^{ad}_{\downarrow}$, the adatom is
magnetized and if they are equal, it isn't magnetized. The zero
temperature occupation number of the adatom level for spin
$\sigma$ is given by
\begin{eqnarray}
n_{\sigma}^{ad}=\int_{-\infty}^{\mu}d\omega
\rho_{\sigma}^{ad}(\omega) ,\label{eq:07}
\end{eqnarray}
where
$\rho_{\sigma}^{ad}(\omega)=-\frac{1}{\pi}ImG^{ad,R}_{\sigma}(\omega)$
and one can use the equation of motion technique to write
$G^{ad,R}_{\sigma}(\omega)$ as
\begin{eqnarray}
G^{ad,R}_{\sigma}(\omega)=[\omega-\varepsilon_{\sigma}-\Sigma^{ad,R}(\omega)+i0^{+}]^{-1}
.\label{eq:08}
\end{eqnarray}
Here $\Sigma^{ad,R}(\omega)$ is the retarded self-energy and it
can be written as
\begin{eqnarray}
\Sigma^{ad,R}(\omega)=V^{2}_{f}G^{0R}_{\sigma}(\omega)=\frac{V^{2}}
{\sqrt{N}}\sum_{\mathbf{q}}G^{0R}_{\sigma}(\mathbf{q},\omega)
,\label{eq:09}
\end{eqnarray}
where $G^{0R}_{\sigma}(\mathbf{q},\omega)$ is the pure Green's
function of biased AA-stacked BLG. After integration over the
momentum we obtained a relation for the Green's functions as
\begin{eqnarray}
G^{0R}_{\sigma}(\omega)&=&-\frac{1}{2D^{2}}[(\omega-\Pi)\frac{\Pi+V}{\Pi}\ln|\frac{D^{2}-(\omega-\Pi)^{2}}
{(\omega-\Pi)^{2}}|]\nonumber\\&+&\frac{1}{2D^{2}}
[(\omega+\Pi)\frac{\Pi-V}{\Pi}\ln|\frac{D^{2}-(\omega+\Pi)^{2}}{(\omega+\Pi)^{2}}|]\nonumber\\
&-&i\frac{\pi}{2D^{2}}[\frac{\Pi+V}{\Pi}|\omega+\Pi|+\frac{\Pi-V}{\Pi}|\omega-\Pi|]
\nonumber\\&\times&\theta(D-|\omega|).\label{eq:11}
\end{eqnarray}
One can solve eqs. (\ref{eq:07})-(\ref{eq:11}) self-consistently
to obtain the occupation number and determine the magnetization of
the adatom. We present our numerical results in the next section.

\section{Numerical results and discussion}
\label{sec:4}

Here we present our numerical results for magnetization of an
adatom adsorbed on AA-stacked BLG. Firstly, we consider dependence
of the adatom magnetic phase diagram on the adatom energy level
for unbiased AA-stacked BLG (the adatom magnetic phase diagram
separates the magnetic and nonmagnetic phase of the adatom).
Secondly, we investigate the effect of the bias voltage on the
magnetization of the adatom.

Unbiased AA-stacked BLG has special DOS, constant DOS (similar to
normal metals) at low energies and linear DOS (similar to
graphene) otherwise. This can lead to different magnetic phase
diagrams for the adatom, depending on the energy level of the
adatom. Figure \ref{f.03} shows our results for plots of the
boundary between the magnetic and nonmagnetic state of an adatom
adsorbed on unbiased AA-stacked BLG, for three different values of
the adatom energy level, $|\varepsilon_{0}|=0.25\gamma$ ,
$|\varepsilon_{0}|\sim \gamma$ and $|\varepsilon_{0}|=2.0\gamma$.
The left panels correspond to $\varepsilon_{0}<0$ and the right
ones correspond to $\varepsilon_{0}>0$. When the energy level of
the adatom approaches to zero (here for example
$|\varepsilon_{0}|=0.25\gamma$), the magnetic boundary becomes
symmetric around the y=0.5 approximately and does not cross the
y=0.0 (or y=1.0). This is similar to the phase diagram of a
magnetic atatom embedded in the normal metals\cite{Anderson}. This
behavior can be understood by this fact that DOS of AA-stacked BLG
around $\omega=0$ is constant and nonzero (fig. \ref{f.02} (b)).
While for $\varepsilon_{0}\sim +\gamma$ ($\varepsilon_{0}\sim
-\gamma$), the magnetic boundary crosses the line y=0.0 (y=1.0)
namely the adatom can be magnetized even when the bare energy
level is above (under) the Fermi level energy. This feature is due
to the large broadening of the adatom energy level. If
$|\varepsilon_{0}|$ increase more, the magnetic boundary becomes
more asymmetric as the magnetic phase diagram of adatom in
graphene. The large broadening of the adatom energy level (even
larger than the broadening of an adatom level with
$|\varepsilon_{0}|\sim \gamma$), which is due to the linear energy
dependence of DOS around the adatom energy level, is origin of
this feature. Furthermore we see that for large
$|\varepsilon_{0}|$ (larger than inter layer hopping) the magnetic
area shrink in the $\pi V_{f}^{2}/DU$ direction, namely the adatom
magnetization is possible only for larger values of U. This is due
to the large amount of the DOS around the adatom energy level
which enhances the hybridization of adatom energy level with
conduction sea and so limits the magnetization of the adatom. We
see that the adatom adsorbed on AA-stacked BLG shows different
magnetic phase diagrams, depending on the energy level of the
adatom, which varies from local moment phase diagram in normal
metals to that in graphene. This is a new aspect not found in
graphene and AB-stacked BLG.

Hereafter, we investigate the effect of an applied bias voltage on
the magnetization of the adatom adsorbed on undoped AA-stacked
BLG. As shown in Fig. \ref{f.02} (b), DOS of AA-stacked BLG can be
manipulate by applying the bias voltage. This allows to control
the hybridization of the adatom energy level with the conduction
sea of AA-stacked BLG and to control the magnetization of the
adatom. This is possible for a wide range of the adatom energy
levels and on site coulomb energies as shown in figs. \ref{f.04}
and \ref{f.05}. In fig. \ref{f.04}, plots of magnetization versus
the bias voltage have been shown for different energy levels. We
see that for a wide range of adatom energy levels, there is a bias
voltage which can be applied to suppress the DOS around the adatom
energy level. This decreases the hybridization of the adatom
energy level with conduction sea and so the magnetization of the
adatom. Fig. \ref{f.05} shows plots of the magnetization of the
adatom for different on site coulomb energies. We see that even at
a small on site coulomb energy about 0.1 eV, the adatom is
magnetized with a large amount of the magnetic moment about 0.7
$\mu_{B}$ and its magnetization can be controlled by the bias
voltage. Controlling the magnetization of the adatom, specially
for a such wide range of the adatom energy levels and the on site
coulomb energies, allows for possibility of using AA-stacked BLG
in spintronic devices.

\section{Summary and conclusion}
\label{sec:5}

In summary, we applied the Anderson impurity model to study
magnetization of an adatom adsorbed on AA-stacked BLG. We did our
investigation for both biased and unbiased AA-stacked BLG. In the
unbiased case, we found different magnetic phase diagrams for the
adatom. We showed that this varies from magnetic phase diagram of
an adatom embedded in normal metals to that for an adatom adsorbed
on graphene surface, depending on the bare energy level of the
adatom. We explained this feature based on the individual energy
dependence of the DOS of AA-stacked BLG and anomalous broadenings
of the adatom energy level. Furthermore we investigated the effect
of an applied bias voltage on DOS of AA-stacked BLG and showed
that one can control the magnetization of the adatom via the bias
voltage for wide range of adatom energy levels and on site coulomb
energies. This allows for application of AA-stacked BLG in
spintronic devices.

%\newpage

\newpage

\begin{figure}
\includegraphics[width=15cm,height=10cm,angle=0]{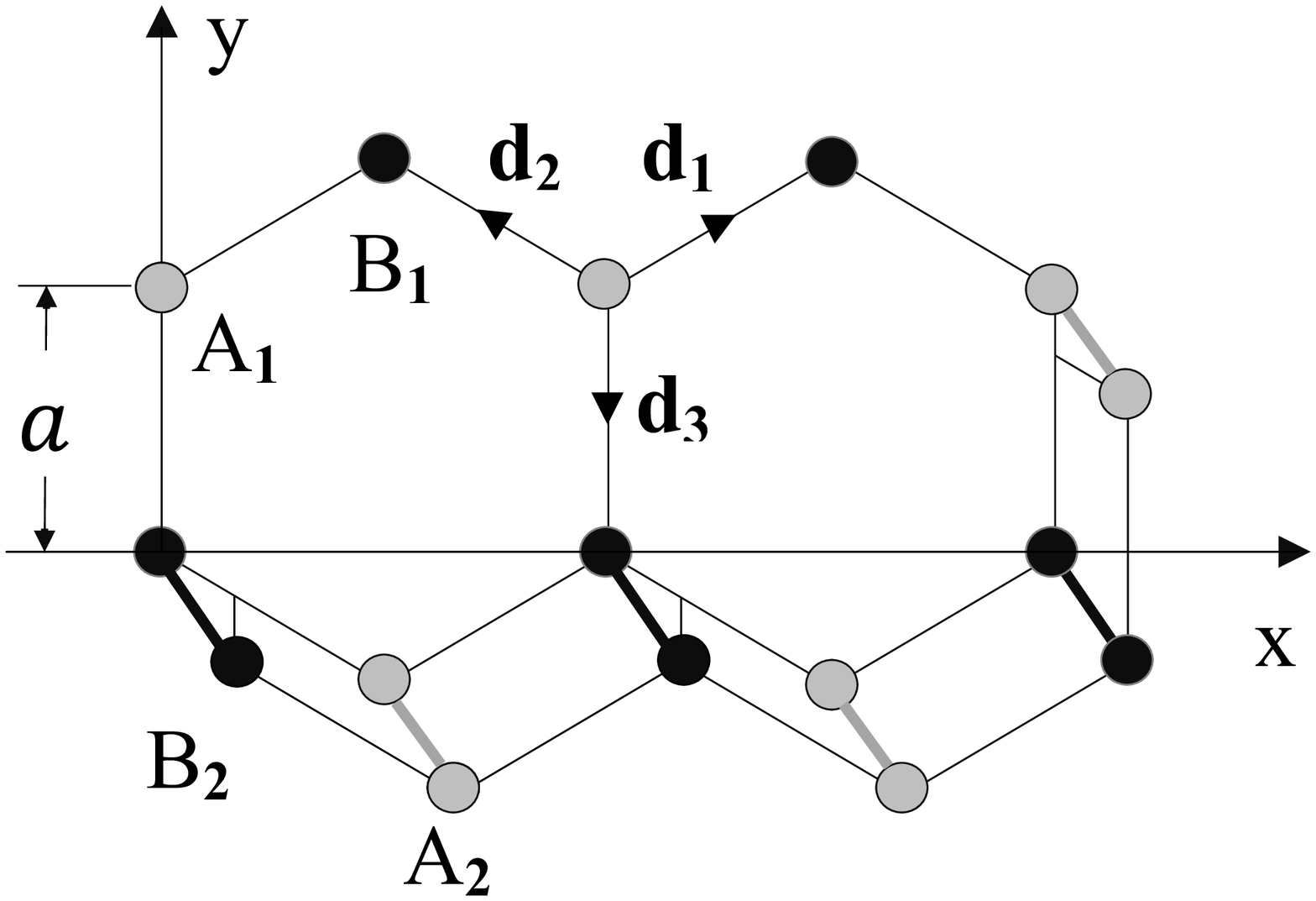}
\caption{Lattice structure of AA-stacked BLG.
$\mathbf{d}_{1}=(a\sqrt{3}/2,a/2)$,
$\mathbf{d}_{2}=(-a\sqrt{3}/2,a/2)$ and $\mathbf{d}_{3}=(0,-a)$
are three vectors that are drown from connects $A_{1}$ sublattice
to its nearest neighbors.} \label{f.01}
 \end{figure}
\begin{figure}
\includegraphics[width=15cm,height=10cm,angle=0]{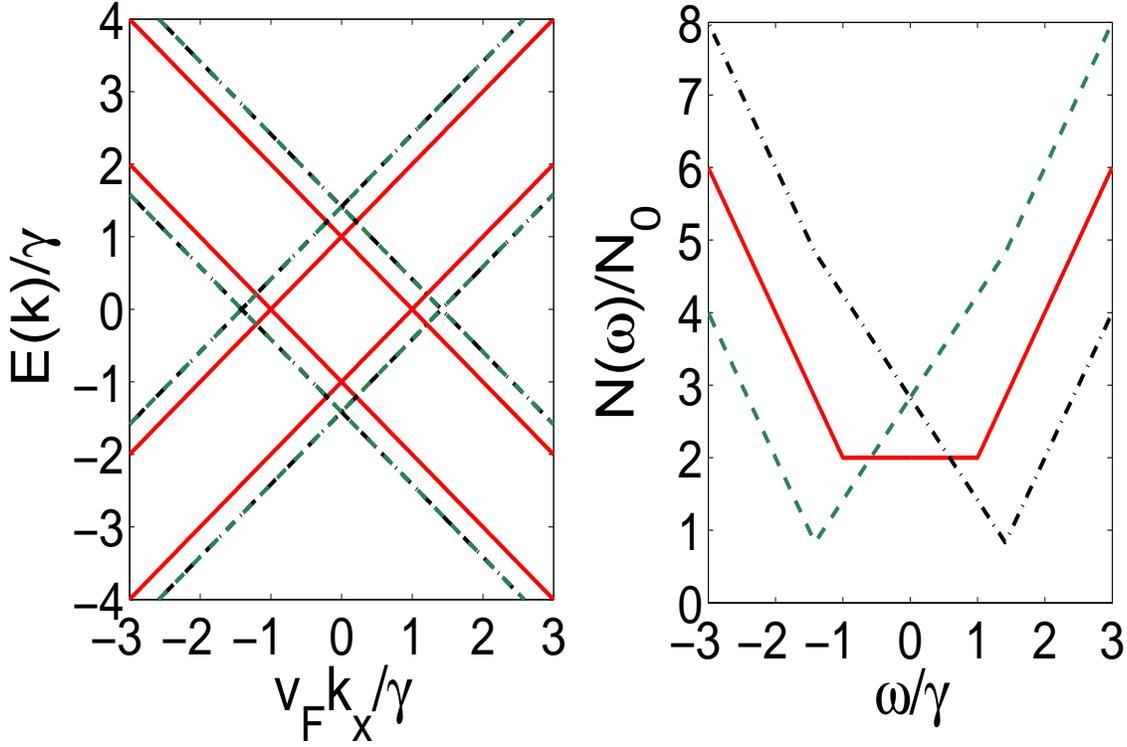}\caption{(a)
shows the low-energy band structure of biased AA-stacked BLG for
three values of the bias voltage $V=0.0$ (solid line), $V=-\gamma$
(dashed line) and $V=+\gamma$ (doted-dashed line) and (b) shows
the DOS on a sublattice of top layer for these values of the bias
voltage where $N_{0}=2\gamma/v_{F}^{2}$.} \label{f.02}
\end{figure}
\begin{figure}
\includegraphics[width=15cm,height=10cm,angle=0]{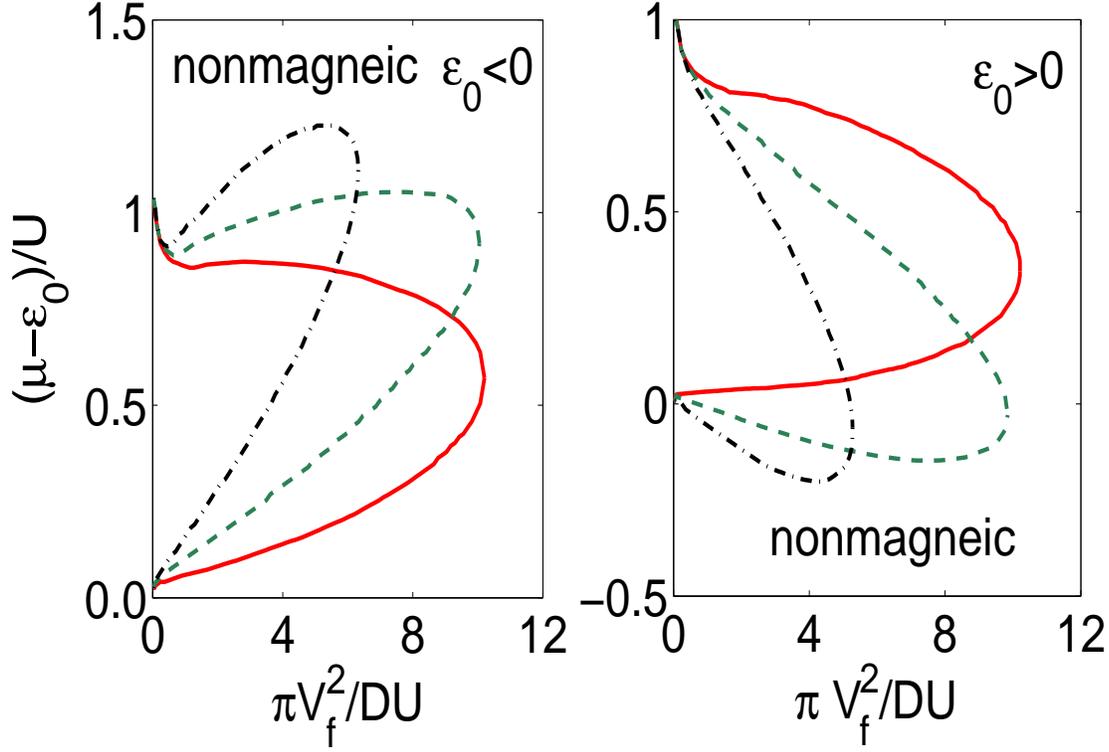}\caption{The
boundary between magnetic and non-magnetic states of an adatom
adsorbed on a sublattice of the unbiased AA-stacked BLG for three
different values of the adatom energy level,
$|\varepsilon_{0}|=0.25\gamma$ (solid curves),
$|\varepsilon_{0}|=1.00 \gamma$(dashed curves) and
$|\varepsilon_{0}|=2.00 \gamma$(doted-dashed curves). The right
panels are for $\varepsilon_{0}<0$ and the left ones for
$\varepsilon_{0}>0$. The other parameters are $\gamma=0.2 eV$,
$V_{f}=1 eV$ and $D\sim 7 eV$.} \label{f.03}
\end{figure}
\begin{figure}
\includegraphics[width=15cm,height=10cm,angle=0]{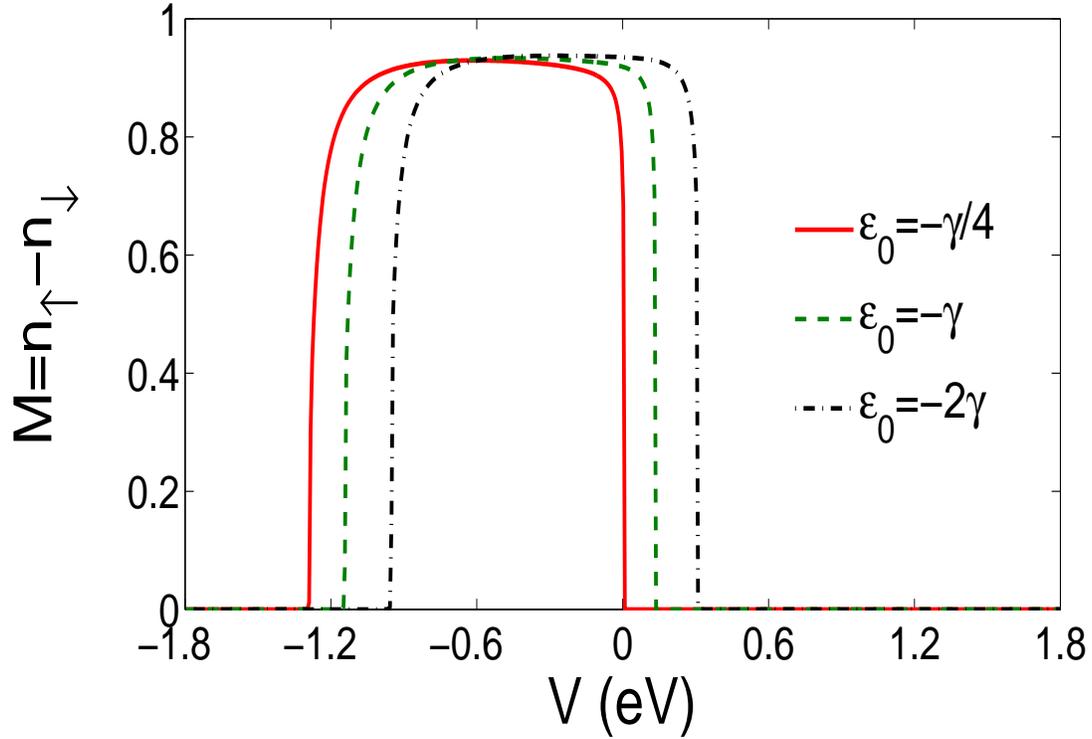}\caption{
Plots of magnetization of the adatom vs. the applied bias voltage
for different values of the adatom energy level,
$\varepsilon_{0}=-\gamma/4$, $\varepsilon_{0}=-\gamma$ and
$\varepsilon_{0}=-2\gamma$. The other parameters are $\gamma=0.2
eV$, $V_{f}=1 eV$, U=1.5 eV and $\mu=0$.} \label{f.04}
\end{figure}
\begin{figure}
\includegraphics[width=15cm,height=10cm,angle=0]{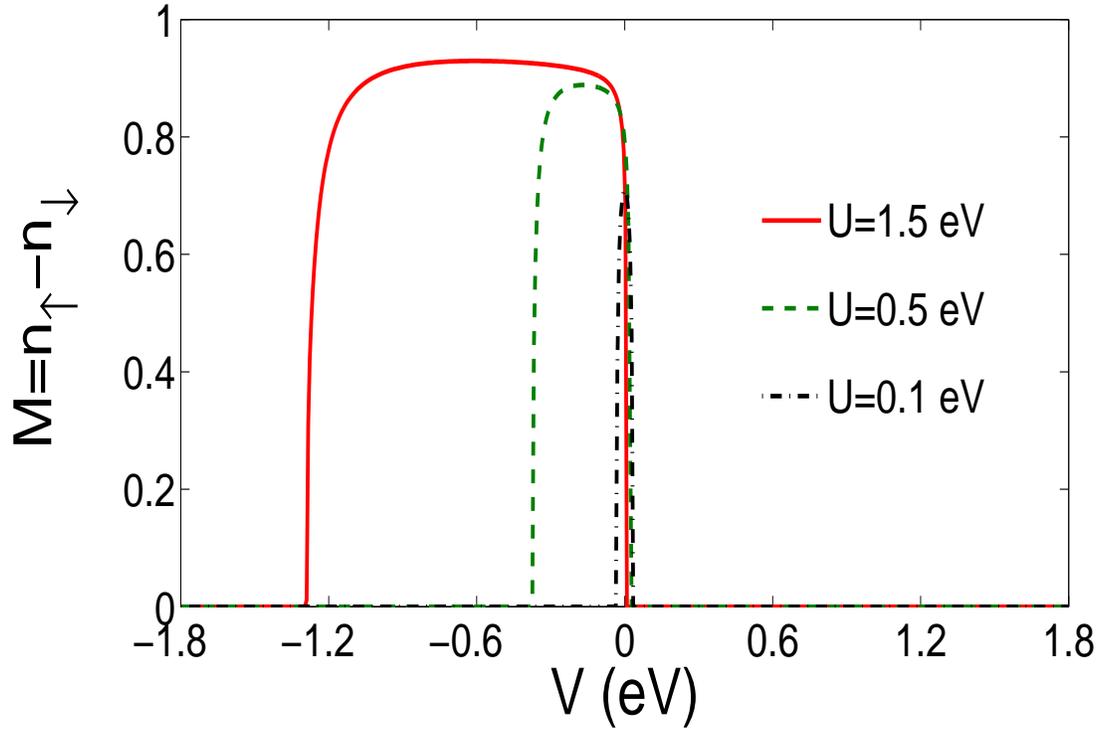}\caption{M(V)
of the adatom for different values of the on site coulomb energy,
U=1.5 eV, U=0.5 eV and U=0.1 eV. The other parameters are
$\gamma=0.2 eV$, $V_{f}=1 eV$, $\varepsilon_{0}=-\gamma/4$ and
$\mu=0$.} \label{f.05}
\end{figure}

\end{document}